\documentclass[iop,revtex4]{emulateapj}
\usepackage{natbib}
\slugcomment{Published in The Astrophysical Journal Letters}

\shorttitle{CO observations of high-z starbursts}
\shortauthors{Silverman et al.}

\begin{document}


\title{A higher efficiency of converting gas to stars pushes galaxies at $z\sim1.6$ well-above the star-forming main sequence} 


\author{J. D. Silverman\altaffilmark{1}, E.~Daddi\altaffilmark{2}, G.~Rodighiero\altaffilmark{3}, W. Rujopakarn\altaffilmark{1,4},  M.~Sargent \altaffilmark{5}, A.~Renzini\altaffilmark{6}, D. Liu\altaffilmark{2}, C. Feruglio\altaffilmark{7}, D. Kashino\altaffilmark{8}, D.~Sanders\altaffilmark{9}, J.~Kartaltepe\altaffilmark{10}, T.~Nagao\altaffilmark{11}, N.~Arimoto\altaffilmark{12}, S. Berta\altaffilmark{13}, M.~B{\'e}thermin\altaffilmark{14}, A.~Koekemoer\altaffilmark{15}, D. Lutz\altaffilmark{13}, G. Magdis\altaffilmark{16,17}, C. Mancini\altaffilmark{6}, M. Onodera\altaffilmark{18}, G. Zamorani\altaffilmark{19}}
\email{john.silverman@ipmu.jp}


\altaffiltext{1}{Kavli Institute for the Physics and Mathematics of the Universe (WPI), The University of Tokyo Institutes for Advanced Study, The University of Tokyo, Kashiwa, Chiba 277-8583, Japan}
\altaffiltext{2}{Laboratoire AIM, CEA/DSM-CNRS-Universite Paris Diderot, Irfu/Service d'Astrophysique, CEA Saclay}
\altaffiltext{3}{Dipartimento di Fisica e Astronomia, Universita di Padova, vicolo Osservatorio, 3, 35122, Padova, Italy}
\altaffiltext{4}{Department of Physics, Faculty of Science, Chulalongkorn University, 254 Phayathai Road, Pathumwan, Bangkok 10330, Thailand}
\altaffiltext{5}{Astronomy Centre, Department of Physics and Astronomy, University of Sussex, Brighton, BN1 9QH, UK}
\altaffiltext{6}{Instituto Nazionale de Astrofisica, Osservatorio Astronomico di Padova, v.co dell'Osservatorio 5, I-35122, Padova, Italy, EU}
\altaffiltext{7}{IRAM - Institut de RadioAstronomie Millim\'etrique, 300 rue de la Piscine, 38406 Saint Martin d'H\`eres, France}
\altaffiltext{8}{Division of Particle and Astrophysical Science, Graduate School of Science, Nagoya University, Nagoya, 464-8602, Japan}
\altaffiltext{9}{Institute for Astronomy, University of Hawaii, 2680 Woddlawn Drive, Honolulu, HI, 96822}
\altaffiltext{10}{National Optical Astronomy Observatory, 950 N. Cherry Ave., Tucson, AZ, 85719}
\altaffiltext{11}{Graduate School of Science and Engineering, Ehime University, 2-5 Bunkyo-cho, Matsuyama 790-8577, Japan}
\altaffiltext{12}{Subaru Telescope, 650 North A'ohoku Place, Hilo, Hawaii, 96720, USA}
\altaffiltext{13}{Max-Planck-Institut f\"ur extraterrestrische Physik, D-84571 Garching, Germany}
\altaffiltext{14}{European Southern Observatory, Karl-Schwarzschild-Str. 2, 85748 Garching, Germany}
\altaffiltext{15}{Space Telescope Science Institute, 3700 San Martin Drive, Baltimore, MD, 21218, USA}
\altaffiltext{16}{Department of Physics, University of Oxford, Keble Road, Oxford OX1 3RH, UK}
\altaffiltext{17}{Institute for Astronomy, Astrophysics, Space Applications and Remote Sensing, National Observatory of Athens, GR-15236 Athens, Greece}
\altaffiltext{18}{Institute of Astronomy, ETH Z\"urich, CH-8093, Z\"urich, Switzerland}
\altaffiltext{19}{INAF Osservatorio Astronomico di Bologna, via Ranzani 1, I-40127, Bologna, Italy}


\begin{abstract}

Local starbursts have a higher efficiency of converting gas into stars, as compared to typical star-forming galaxies at a given stellar mass, possibly indicative of different modes of star formation.  With the peak epoch of galaxy formation occurring at $z>1$, it remains to be established whether such an efficient mode of star formation is occurring at high-redshift.  To address this issue, we measure the molecular gas content of seven high-redshift ($z\sim1.6$) starburst galaxies with the Atacama Large (sub-)Millimeter Array and IRAM/Plateau de Bure Interferometer.  Our targets are selected from the sample of Herschel far-infrared detected galaxies having star formation rates ($\sim$300-800 M$_{\odot}$ yr$^{-1}$) elevated ($\gtrsim4\times$) above the star-forming `main sequence', and included in the FMOS-COSMOS near-infrared spectroscopic survey of star-forming galaxies at $z\sim1.6$ with Subaru.  We detect CO emission in all cases at high levels of significance, indicative of high gas fractions ($\sim$30-50\%). Even more compelling, we firmly establish with a clean and systematic selection that starbursts, identified as main-sequence outliers, at high redshift generally have a lower ratio of CO to total infrared luminosity as compared to typical main-sequence star-forming galaxies, although with a smaller offset than expected based on past studies of local starbursts.  We put forward a hypothesis that there exists a continuous increase in star formation efficiency with elevation from the main sequence with galaxy mergers as a possible physical driver.  Along with a heightened star formation efficiency, our high-redshift sample is similar in other respects to local starbursts such as being metal rich and having a higher ionization state of the interstellar medium. 
          
\end{abstract}


%% subject header.  Objects should be in the appropriate "individual"
%% headers (e.g. quasars: individual, stars: individual, etc.) with the
%% additional provision that the total number of headers, including each
%% individual object, not exceed six.  The \objectname{} macro, and its
%% alias \object{}, is used to mark each object.  The macro takes the object
%% name as its primary argument.  This name will appear in the paper
%% and serve as the link's anchor in the electronic edition if the name
%% is recognized by the data centers.  The macro also takes an optional
%% argument in parentheses in cases where the data center identification
%% differs from what is to be printed in the paper.

\keywords{galaxies: ISM --- galaxies: high-redshift --- galaxies: starburst --- galaxies: star formation}


\section{Introduction}

It has recently come to light that the growth of galaxies may be less erratic than previously thought.  The existence of a tight relation between the stellar mass ($M_*$) and star formation rate (SFR) of star-forming galaxies, termed the `main sequence' (MS, \citealt[][]{ Noeske2007,Daddi2007,Elbaz2007,Whitaker2012,Kashino2013,Speagle2014}) is indicative of a quasi-equilibrium being maintained between gas supply, SFR, and gas expulsion from galaxies \citep[e.g.,][]{Bouche2010,Lilly2013}.  This SFR-M$_*$ relation evolves strongly with look-back time, paralleling the global evolution of the cosmic SFR density from the present to $z\sim2$ \citep{Madau2014}.  It is becoming clear that such a decline in SFR is in response to diminishing gas reservoirs \citep[e.g.,][]{Tacconi2013,Scoville2014,Santini2014}.  The efficiency of star formation ($SFE\equiv SFR/M_{gas}$), i.e., the efficiency  at which gas is being converted to stars, is remarkably similar across a wide range of cosmic time within the global star-forming population, usually with gas mass $M_{gas}$ inferred from the CO molecular line luminosity $L^\prime_{\rm CO}$  \citep[e.g.,][]{Tacconi2010,Tacconi2013,Daddi2010b}.

However, outliers from  the MS are known to exist, with very high specific SFR (sSFR) compared to normal MS galaxies, such as local ultra-luminous infrared galaxies (ULIRG,  \citealt{Sanders1996}). At $z\sim 2$, these MS outliers represent $ \sim 2\%$ of the star-forming population and contribute only a moderate fraction ($\sim 10\%$) of the global co-moving SFR-density \citep{Rodighiero2011}. Yet,  they are still likely to play an important role in galaxy evolution, as a high fraction of galaxies may experience such a star-bursting event.  Therefore, it is important to understand the physical conditions that can lead to such a surge in SFR.

Local starbursts have a higher SFE compared to MS galaxies.  Estimates of their molecular gas content from their CO luminosity indicate that a given gas mass is capable of producing higher  SFRs \citep{Solomon1997}, and a bi-modal distribution of SFEs  has been proposed for the general MS and outlier population \citep{Daddi2010b, Genzel2010}.  While there are uncertainties with respect to the appropriate conversion of $L^\prime_{\rm CO}$ to H$_2$ gas mass ($\alpha_{\rm CO}$), an offset with respect to MS galaxies is already apparent when considering the observed quantities $L^\prime_{\rm CO}$ and total infrared luminosity ($L_{\rm TIR}$).  The physical mechanism responsible for this enhanced SFE is not entirely clear and one cannot exclude that  sample selection may be responsible for the lack of galaxies with intermediate SFEs.  While CO detections have been obtained for high-redshift submm-selected galaxies \citep[SMGs; e.g.,][]{Greve2005,Bothwell2013} and star-forming radio galaxies \citep[SFRG; ][]{Casey2011}, these samples do not  cleanly distinguish starburst outliers from the MS population and submm selected samples could be biased towards galaxies with lower dust temperatures \citep{Magnelli2014}.  Therefore, it is imperative to study the CO emission of starbursts securely elevated above the MS at their respective stellar mass  and near  the peak epoch of star formation ($z\sim2$).  \citet{Rodighiero2011} achieve such a clear distinction between MS and MS-outlier galaxies using large photometric samples with multi-wavelength coverage, including the bulk of the FIR emission. Our targets then result in a homogenous sample that does not have contaminants such as those inherent with other selections.  

\begin{figure}
\epsscale{1.1}
\plotone{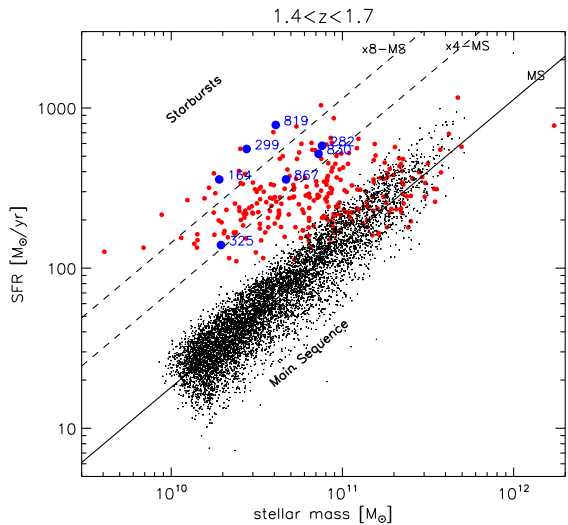}
\caption{SFR versus M$_*$ for galaxies with $1.4<z<1.7$ in COSMOS.  Galaxies with CO observations are shown in blue.  For reference, we plot star-forming galaxies that denote the MS (small dots).  $Herschel$-detected galaxies include our sample in blue and those shown in red.}
\label{fig:sample}
\end{figure}

To address this issue, we report on interferometric observations of the molecular transition $^{12}$CO 2-1 and $^{12}$CO 3-2 for a sample of seven MS outliers using both the Atacama Large Millimeter Array (ALMA) and IRAM Plateau de Bure Interferometer.  Our sample is extracted from the 246 MS outliers (with sSFRs $\gtrsim4\times$ above the mean of the MS; Figure~\ref{fig:sample}) at $1.4<z<2.5$ identified with Herschel observations over the COSMOS field \citep{Rodighiero2011} that effectively cover the peak FIR emission. Among these Herschel-detected galaxies, those with photometric redshifts between 1.4 and 1.7 are observed through our Subaru Intensive Project \citep{Silverman2015} with the FMOS near-IR multi-object spectrograph.  These spectra provide a detection of key diagnostic emission lines (i.e., H$\alpha$, H$\beta$, [NII]$\lambda6585$, [OIII]$\lambda$5008) used to measure accurate spectroscopic redshifts, SFRs, and metallicities.  Six targets, labeled in Figure~\ref{fig:sample}, are extracted from this sample with an additional galaxy at a slightly higher redshift (PACS-282; $z=2.19$) where the redshift comes from the zCOSMOS-Deep program \citep{Lilly2007}.

Our confidence in these galaxies being `outliers' in the SFR$-M_*$  plane is based on the exquisite multi-wavelength coverage  of the COSMOS field.  Stellar masses are  derived from SED-fitting using Hyperz with stellar population synthesis models \citep{Bruzual2003} at the respective spectroscopic redshift.  The de-blending of detections in Herschel (or 70$\micron$ {\it Spitzer}) images relies upon $Spitzer$ MIPS 24$\micron$ priors.  SFRs are derived from L$_{TIR}$, an integral of the best-fit model SED \citep{Draine2007} from 8  - 1000 $\mu$m using the Spitzer (24 $\mu$m), Herschel PACS (100 and 160 $\mu$m) and SPIRE (250, 350 and 500 $\mu$m) bands, thus accounting for the obscured star formation as in \citet{Magdis2012}.  Moreover, four of the seven galaxies have 1.4 GHz  radio detections at $>5\sigma$ level \citep{Schinnerer2010} and radio-based SFRs are consistent with being above the MS.  The presence of strong AGN within our sample is ruled out by the lack of individual X-ray detections by $Chandra$.  Even if an obscured AGN were present, the contribution to L$_{TIR}$ would be negligible \citep{Rodighiero2011,Pozzi2012}.  Throughout this work, we assume $H_0=70 $ km s$^{-1}$ Mpc$^{-1}$, $\Omega_{\Lambda}=0.7$, $\Omega_{\rm{M}}=0.3$, and use a Salpeter IMF for SFRs and stellar masses.
    
\section{Observations of CO emission at high-z}

\begin{figure*}
\epsscale{1.1}
\plotone{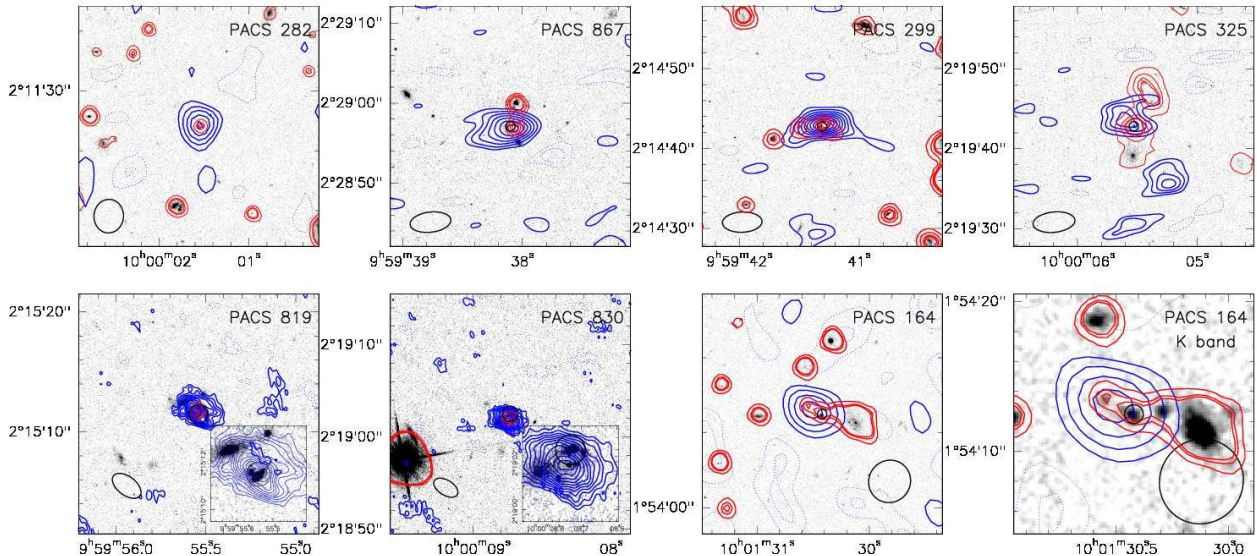}
\caption{HST/ACS F814W images of our seven starbursts.  CO emission is shown as solid (dashed) blue contours in positive (negative) steps of $1\sigma$ starting at 2 (-2) times the noise level (where $\sigma_{rms}$=0.17, 0.12, 0.11, 0.05, 0.12, 0.14 and 0.11 mJy beam$^{-1}$ starting from left to right with 282 and then 819).  $Spitzer$/IRAC 3.6$\mu$m detections are marked with red contours to illustrate the association of peak CO emission with their IR counterparts.  CO beam size is indicated in the lower left corner.  Small black circles show the placement of FMOS fibers. For PACS-164 the $K$-band image is shown to further illustrate the heavily-obscured nature of the CO-emitting region.  Angular sizes of the cutouts varies.}   
\label{fig:maps}
\end{figure*}

\begin{deluxetable*}{llllllllllll}
\tabletypesize{\scriptsize}
\tablecaption{Sample data and CO measurements\tablenotemark{a} \label{tab:sample}}
\tablehead{\colhead{ID}&\colhead{RA}&\colhead{Dec}&\colhead{$z_{CO}$\tablenotemark{b}}&\colhead{$z_{spec}$\tablenotemark{c}}&\colhead{log M$_{*}$\tablenotemark{d}}&\colhead{SFR\tablenotemark{e}}&\colhead{I$_{CO}$\tablenotemark{f}}&\colhead{$L_{CO}^{\prime}$\tablenotemark{g}}&\colhead{$\Delta$v\tablenotemark{h}}&\colhead{CO}&\colhead{Gas\tablenotemark{i}}\\
&\colhead{(CO)}&\colhead{(CO)}&&&\colhead{M$_{\odot}$}&&\colhead{Jy km s$^{-1}$}&&&\colhead{size ($\arcsec$)}&\colhead{fraction}}
\startdata
299&09:59:41.31&02:14:42.91&1.6483&1.6467&10.44&554$\pm40$&0.67$\pm$0.08&10.44&590&$<$2.4&0.52\\
325&10:00:05.47&02:19:42.61&&1.6557 &10.29&139$\pm10$&0.28$\pm$0.06&10.07&764&&0.32\\ 
819&09:59:55.55&02:15:11.70&1.4451&1.4449&10.61&783$\pm18$&1.10$\pm$0.07&10.55&592&0.34$\pm$0.08&0.34\\
830&10:00:08.75&02:19:01.90&1.4631&1.4610&10.86&517$\pm12$&1.18$\pm0.10$&10.59&436&0.97$\pm$0.17&0.46\\ 
867& 09:59:38.12&02:28:56.56&1.5656&1.5673&10.67&358$\pm34$&0.46$\pm0.04$&10.24&472&$<$1.5&0.29\\
282&10:00:01.54&02:11:24.27&2.1869&2.1924&10.88& 581$\pm42$&0.75$\pm$0.12&10.44&660&$<$3.4&0.28\\
164&10:01:30.53&01:54:12.96&1.6481&1.6489&$>$10.28&358$\pm8$&0.61$\pm$0.11&10.40&894&$<$4.8&0.52
\enddata
\tablenotetext{a}{The first five targets are observed with ALMA while the remaining two with PdBI.  }
\tablenotetext{b}{$\sigma_z$ = 0.0003-0.0004.  An accurate CO centroid was not obtained for 325 thus the H$\alpha$ redshift was used for the CO flux measurement.}
\tablenotetext{c}{Spectroscopic redshifts are based on H$\alpha$ with the exception of $\#$282 and have errors $\sigma_{\Delta z/(1+z)}=1.8\times10^{-4}$.}
\tablenotetext{d}{$\sigma_{\rm M}\sim0.07$ dex \citep{Ilbert2015}}
\tablenotetext{e}{Units of M$_{\odot}$ yr$^{-1}$.}
\tablenotetext{f}{All are based on the CO 2-1 transition with the exception of 282 (3-2).}
\tablenotetext{g}{{\rm Log} base 10; units of K km s$^{-1}$ pc$^2$}
\tablenotetext{h}{Full width of the CO line at zero intensity in units of velocity (km s$^{-1}$).}
\tablenotetext{i}{$f_{gas}= M_{gas} / (M_{*} + M_{gas}$)}
\end{deluxetable*}

ALMA observations of five galaxies were carried out in Cycle 1 (Project 2012.1.00952.S) using 25-30 12m antennas and the band 3 receiver tuned to intermediate frequencies between 86.395 - 94.888 GHz with a resolution of 488 kHz and spectral bandpass of 1.875 GHz.  Two `scheduling blocks'  (SB1 and SB2) were defined to accommodate the requirements to detect CO 2-1 within a single sideband for each of our targets (SB1: PACS-819, 830; SB2: PACS-325, 299 and 867).  On-source exposure times were 32 minutes (819 and 830), 2.3 hours (299 and 325) and 3.8 hours (867) that met our request based on predictions of CO emission from their IR luminosity and an assumed ratio L$_{\rm CO}$/L$_{\rm TIR}$ (3$\times$ lower than MS galaxies).  The antenna configurations had baselines up to $\sim$400m ($\sim$200m) for SB1 (SB2) thus resulting in a beam size (FWHM) of $1.3\arcsec \times 1.0\arcsec$ ($4.5\arcsec \times 2.0\arcsec$; $5.4\arcsec \times 2.7\arcsec$ for PACS 325 only).  The images of PACS-819 (830) have a flux sensitivity level of 0.12 (0.10) mJy beam$^{-1}$ over a bandwidth of 400 km s$^{-1}$ while the deeper observations reach 0.073 - 0.14 mJy beam$^{-1}$ for the same spectral window.  Standard targets were used for flux (Mars, Ganymede), bandpass, and phase calibration.
      
We measure the CO flux by fitting the data in UV space with the GILDAS task `uvfit' available in the MAPPING package.  The fit was performed using a point source model or a circular Gaussian for resolved sources that returns the centroid, de-convolved source size, and integrated flux. We find good agreement between fluxes returned from GILDAS with those using `imfit' in the image plane with CASA.  Additionally, we resolve the emission for PACS- 819 and 830; therefore, a measure of the source extent is reported.  

To increase the sample size, interferometric measurements were obtained with IRAM/PdBI for PACS-282 (CO 3-2) and PACS-164 (CO 2-1).  For PACS-282 (PACS-164), an integration time of 6.2 (12.9) hours, using the 3mm band in configuration D, was achieved that resulted in a limiting 1$\sigma$ sensitivity level of 0.14 (0.11) mJy beam$^{-1}$ for a bandwidth of 240 (180) MHz.  CO fluxes are measured with GILDAS.  In Table~\ref{tab:sample}, we list the source properties and CO measurements. 

\begin{figure*}
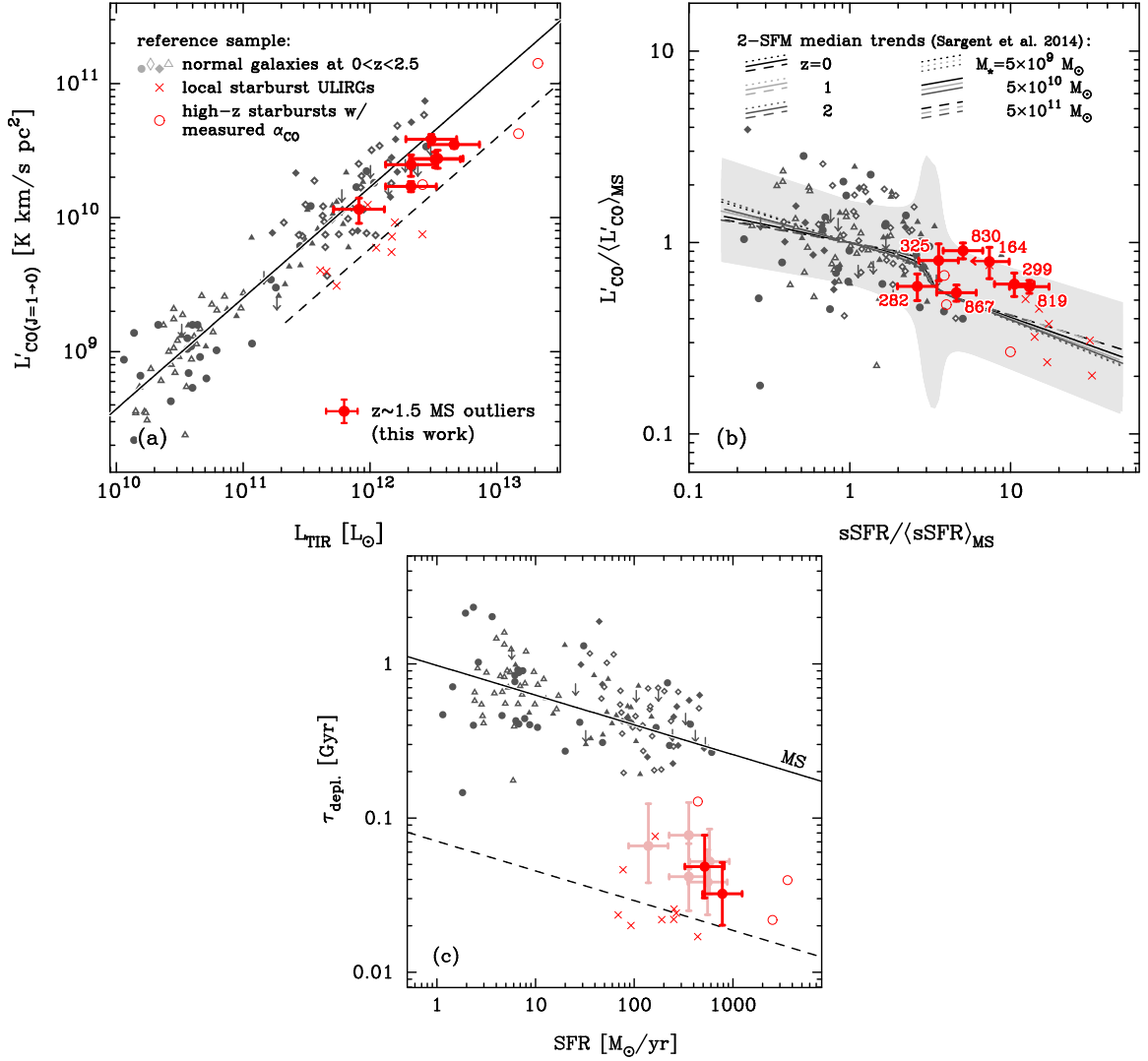

\epsscale{1.0}
\plotone{f3a.ps}
\epsscale{0.45}
\plotone{f3b.ps}
\caption{$(a)$ CO luminosity versus total infrared luminosity.  $(b)$  CO luminosity and specific star formation rate with each normalized to the typical values of MS galaxies.  Empirical model curves and 1$\sigma$ errors (grey region) are described in \citet{Sargent2014}. $(c)$ Gas depletion time ($\tau_{del}\equiv SFE^{-1}$; units Gyr) versus SFR.  Colored symbols with error bars show our starburst sample with two (819 and 830) in bright red having dynamical mass measurements.  Grey symbols represent published samples with $\alpha_{\rm CO}$ estimates as compiled in \citet{Sargent2014} and shown in all panels.}
\label{fig:sf_eff}
\end{figure*}

\section{Results}

We detect CO emission in all seven targeted galaxies with integrated flux densities ($I_{CO}$) ranging from 0.28 to 1.18 Jy km s$^{-1}$.  All but one have a high level of significance ($>5\sigma$) and those resulting from ALMA observations are above the 8$\sigma$ level with the exception of PACS-325 ($S/N=4.7$).  In Figure~\ref{fig:maps}, we display the optical HST/ACS F814W image cutouts \citep{Koekemoer2007} with CO and 3.6$\mu$m emission overlaid as contours.  Using a large beam for the majority of the sample, the CO emission is essentially unresolved with the exception of two observations taken at higher resolution ($\sim1\arcsec$).  PACS-819 has half of the emission coming from a region of 3.2$\pm$0.8 kpc, while PACS-830 is more extended (9.0$\pm$1.6 kpc).  This may suggest that there is diversity in the size distribution of molecular gas at high redshift \citep[e.g., ][]{Ivison2011,Riechers2011,Hodge2012}, dissimilar to local starbursts \citep[e.g.,][]{Scoville1989}.  

Upon close inspection of the maps in Figure~\ref{fig:maps}, it is evident that the centroid of the CO emission is not always coincident  with the brightest regions of UV emission as seen in the HST/ACS image (i.e., PACS-164, PACS-830, PACS-867).  This is likely evidence that a fair fraction of the star formation is obscured as supported by the improved alignment of the CO emission with the peak infrared emission detected by $Spitzer$ in both the IRAC and MIPS channels, and clear association with highly significant radio emission at 1.4 GHz \citep{Schinnerer2010}.    

We note that the Herschel far-IR (and possibly the CO) emission of PACS-164 is likely the sum of two separate components seen in the IRAC image.  The peak of the CO emission, detected by IRAM, is located somewhat between two IR sources, as made particularly clear by the $K$-band image also shown in Figure~\ref{fig:maps}, and slightly elongated possibly indicating a contribution from both sources.  The FMOS spectrum that provides the redshift ($z=1.650$) refers to  the western IR component  that is co-spatial with the UV-bright source seen in the HST image.  With a redshift of the CO emission ($z=1.647$) very close to that of the FMOS source, this system appears to be in the early stages of a merger given the projected separation of 18.7 kpc.  The lower limit for the stellar mass of the system as reported in Table 1 refers to the mass of the object observed with FMOS, and encircled in Figure 2.

\subsection{CO-to-IR ratio}

Our primary interest is to determine whether high-redshift starbursts convert gas into stars with a higher efficiency than MS galaxies at these epochs. To begin with, we use the observed CO to IR luminosity ratio $L^\prime_{\rm CO}/L_{\rm TIR}$  rather than derived quantities, e.g. the gas mass, that depends on the CO-to-H$_2$ conversion factor ($\alpha_{\rm CO}$), or gas mass surface density, which would further require CO-size information for which we only have for 2/7 galaxies in our sample.  However, we note that $L^\prime_{\rm CO}/L_{\rm TIR}\propto M_{\rm gas}/$SFR (modulo $\alpha_{CO}$), hence this luminosity ratio is a fair proxy for the gas depletion time which in turn is the inverse of the SFE. 

In Figure~\ref{fig:sf_eff}$a$, we plot $L^\prime_{\rm CO}$ as a function of $L_{TIR}$, indicative of the obscured SFR, and include both low- and high-redshift galaxies with measurements available in the literature and compiled in \citet{Sargent2014}.  All line luminosities are converted to CO (1-0) using values of 0.85 and 0.7, respectively for CO (2-1) and CO (3-2) (Daddi et al. 2015).  Different excitation corrections than those applied here would not impact our results since changes would either be insignificant (5 - 10$\%$ if CO (2-1) transitions were generally more excited than assumed here) or affect only one of our sources (PACS 282, for which applying the CO(3-2)/CO(1-0) ratio of \citet{Bothwell2013} would result in a 35$\%$ difference).

All galaxies in our sample have $L^\prime_{\rm CO}/L_{\rm TIR}$ below the well-defined correlation for MS galaxies.  These observations indicate the existence of an offset in the $L^\prime_{\rm CO}/L_{\rm TIR}$  ratio for high-redshift starbursts as found by \citet{Solomon1997} for local ULIRGs.  In Figure~\ref{fig:sf_eff}$b$, we plot these quantities, normalized to the mean value of galaxies on the MS.  In addition, on the abscissa we replace $L_{\rm TIR}$  with its implied  specific SFR  (sSFR=SFR/$M_*$), also normalized to the MS value.  We measure a median value of $L^\prime_{\rm CO}/<L^\prime_{\rm CO}>_{\rm MS}=0.60\pm0.04$ for our sample.  This represents an offset ($1.7\pm0.1\times$) from the MS that is smaller than in local samples of starbursts ($\sim3\times$; dashed line in Fig.~\ref{fig:sf_eff}{\it a}).  Along these lines, the starbursts in our sample have higher $L^\prime_{CO}$ than expected by $\sim0.2$ dex (given their excess sSFR relative to MS galaxies) based on the empirical model of \citet{Sargent2014}, shown in Figure~\ref{fig:sf_eff}$b$.

These results may be indicative of a more continuous range in SFE, following the trend for MS galaxies (Fig.~\ref{fig:sf_eff}$b$) and opposed to the notion of a second mode of star-formation operating at higher efficiency distinct from MS galaxies \citep{Daddi2010b, Genzel2010, Genzel2015}.  However, any statement on differences in SFE requires a factor $\alpha_{\rm CO}$ as further addressed below.  Even with continuity in these parameters, there could still exist an underlying bi-modal physical framework for star formation as demonstrated with empirical models shown in Figure~\ref{fig:sf_eff}$b$ and fully explained in \citet{Sargent2014}.       

\subsection{Gas masses, $\alpha_{\rm CO}$, and SFE}

With PACS-819 and PACS-830 having marginally resolved CO emission, we can estimate the total gas mass using the dynamical mass method \citep[e.g.,][]{Tan2014}.  The dynamical mass ($M_{\rm dyn}$) within the half-light radius ($r_{\rm 1/2}$) is given for the spherically symmetric  case by:
\begin{equation}
M_{\rm dyn}(r<r_{\rm 1/2}) \simeq \frac{5 \sigma^2 r_{\rm 1/2}}{G}
\label{eq1}
\end{equation}

\noindent where $\sigma$= $\Delta \upsilon_{\rm CO}$/2.35 is the velocity dispersion, $\Delta \upsilon_{\rm CO}$ is the line width - FWHM, and $G$ is the gravitational constant.  The gas mass is then derived by subtracting from the dynamical mass half of the stellar mass and a dark matter component ($M_{\rm DM}=0.25\times$ $M_{dyn}$; \citealt{Daddi2010a}) as follows.     
\begin{equation}
M_{\rm dyn}=0.5\times(M_*+M_{\rm gas})+M_{\rm DM}(r<r_{\rm 1/2}).
\label{eq2}
\end{equation}

\noindent With this method, the gas mass for PACS-819 (PACS-830) is estimated to be $2.0\pm1.2 \; (6.2\pm2.2)\times10^{10}$ M$_{\odot}$, based on their half-light radii, respectively $1.4\pm0.3~{\rm kpc}$ ($4.1\pm0.7~{\rm kpc}$) and velocity dispersion $\sigma$, respectively $168.0\pm20.2$ km s$^{-1}$ ($148.7\pm 20.8$ km s$^{-1}$), that corresponds to a gas fraction $f_{\rm gas}=M_{\rm gas} / (M_{\rm gas} + M_*$) of 0.34 (0.46).  Based on these values, we estimate  $\alpha_{\rm CO}=0.6\pm0.3$  for PACS-819 and $1.6\pm0.6$ for PACS-830, both not far from those found for the high-redshift submillimeter galaxies in the proto-cluster GN20 \citep{Tan2014} and consistent with  lower conversion factors for galaxies in mergers \citep{Narayanan2011} as compared to isolated disk galaxies (see also \citealt{Genzel2015}).

We roughly estimate the H$_2$ gas masses from the CO line luminosity for the galaxies with unresolved CO emission using the average ($\alpha_{\rm CO}=1.1$ M$_{\odot}$ / K km s$^{-1}$ pc$^{-2}$) value of the two estimates of $\alpha_{\rm CO}$ given above which is in very close agreement to that found for local nuclear starbursts \citep[$\alpha_{\rm CO}=0.8$;][]{Downes1998}.  Based on this conversion factor and keeping in mind the uncertainty, the gas masses are in the range $\sim (1.3-6.2)\times10^{10}$ M$_{\odot}$.  This results in gas fractions $f_{\rm gas}=M_{\rm gas} / (M_{\rm gas} + M_{*}$) between 28 and 52\% (see Table~\ref{tab:sample} for individual estimates), similar to those of MS galaxies at these redshifts \citep{Tacconi2013,Bethermin2015} and in line with cosmological galaxy formation models \citep{Narayanan2012}.      

Based on these results, an increase in the SFE is likely responsible for their elevation above the MS.  In Figure~\ref{fig:sf_eff}$c$, we illustrate this scenario by plotting the gas depletion timescale ($\tau_{del}\equiv SFE^{-1}$; units Gyr) as a function of SFR.  As described above, the lower CO luminosities at a given L$_{TIR}$, compared to MS galaxies, are indicative of a lower gas mass hence shorter gas depletion timescales and higher SFEs, in agreement with previous studies \citep{Genzel2010,Daddi2010b}.  Despite the remaining uncertainties on the assumptions used to derive H$_2$ masses, we see evidence that starbursts in our sample have a smaller contrast (i.e., less extreme) in SFE, relative to MS galaxies, as compared to local ULIRGs and the strongest starbursts at high-z.

\begin{figure}
\epsscale{1.1}
\plotone{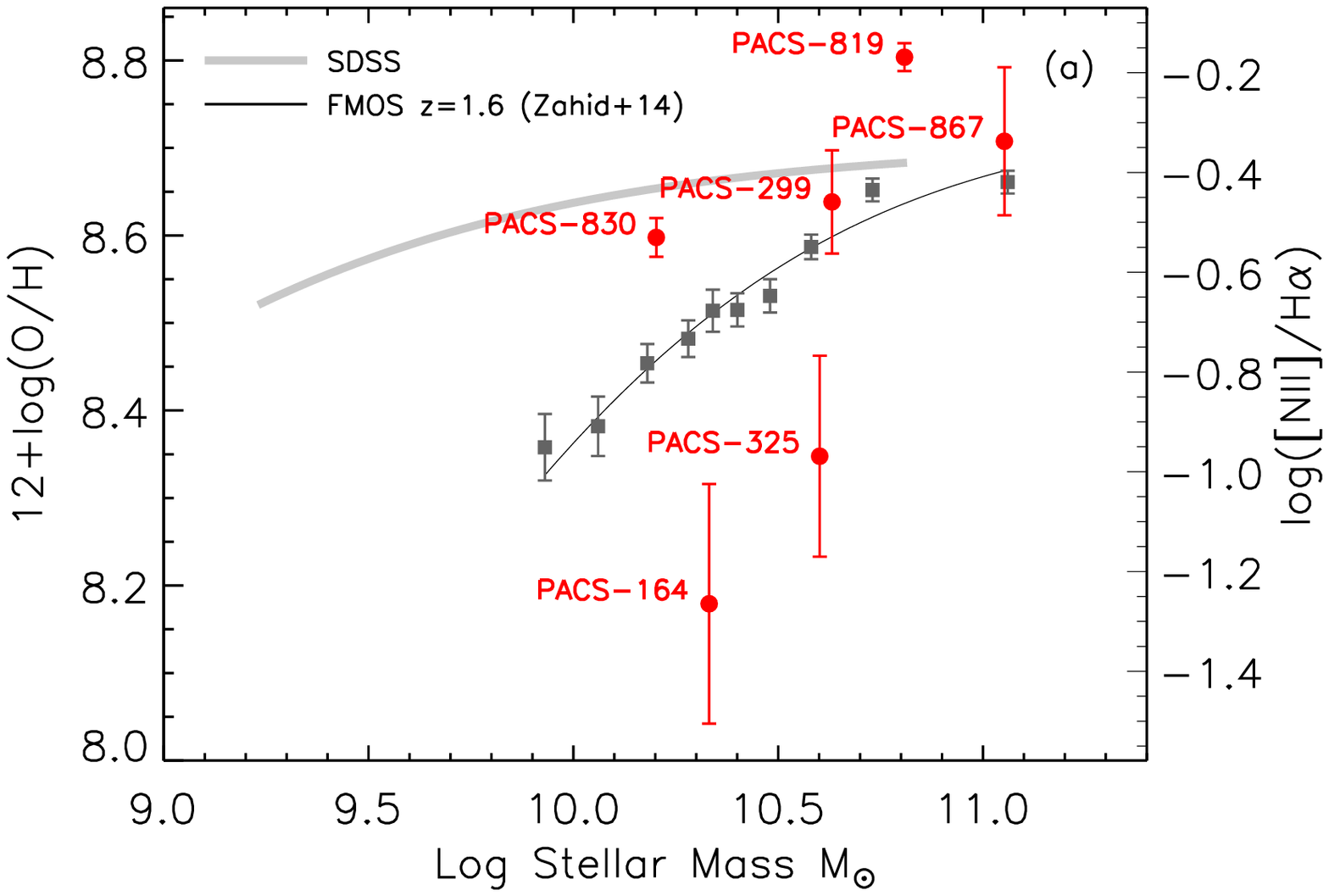}
\epsscale{1}
\plotone{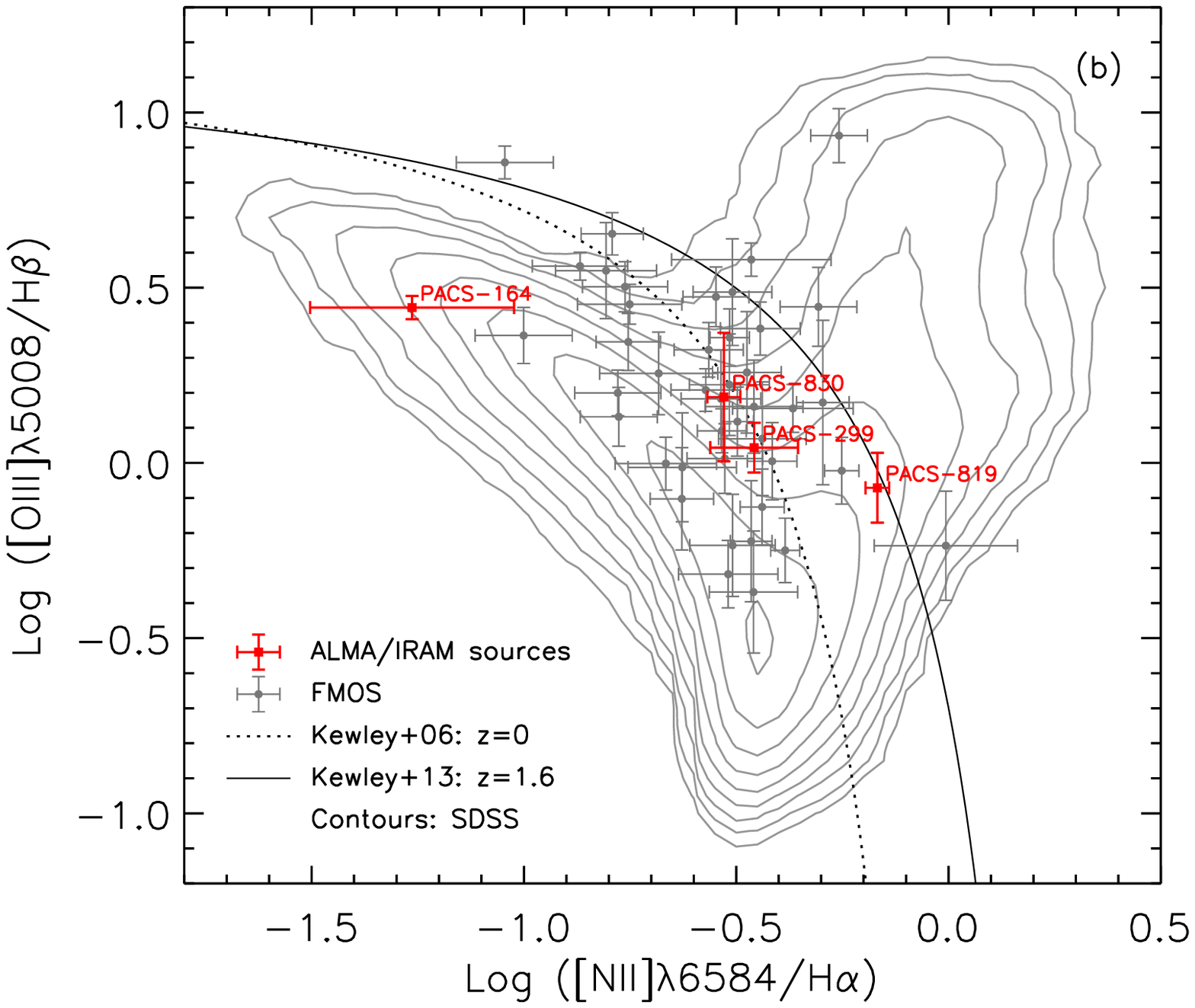}

\caption{Rest-frame emission line ratios from FMOS-COSMOS:  $(a)$ Mass - metallicity relation of our high-z starburst sample (red points).  For comparison, we plot the M-Z relation from SDSS and FMOS at $z\sim1.6$ \citep{Zahid2014}. $(b)$ BPT diagram at $z\sim1.6$ \citep{Kartaltepe2015} with our CO sample as indicated.  Individual measurements, shown in grey, represent galaxies from our larger sBzK sample \citep{Silverman2015}.  Contours denote the region spanned by low-redshift galaxies in SDSS.  Theoretical curves separating star-forming galaxies and AGN (that evolve with redshift) are given.  All data points have $\pm1\sigma$ errors.}
\label{fig:emlines}
\end{figure}

As a final check, we can assess whether the values of $\alpha_{\rm CO}$ derived above are appropriate for our starbursts by using the metallicity of the ISM as a proxy, given that it is well established that $\alpha_{\rm CO}$ is anti-correlated with metallicity \citep[e.g.,][]{Arimoto1996,Schruba2012,Genzel2012}.  In Figure~\ref{fig:emlines}$a$, we show the mass-metallicity relation at $z\sim1.6$ \citep{Zahid2014} based on [NII]$\lambda6585$/H$\alpha$, with our sample shown in red.  We find that 4 out of 6 galaxies (having errors $<\pm0.2$ on the ratio) have high metallicities, similar to those of local massive galaxies.  Therefore, our derived values of $\alpha_{\rm CO}$ are consistent with local galaxies of a similar mass and metallicity, although with the caveat that we are not certain whether the metallicity of the line-emitting gas is representative of the molecular gas producing the CO emission.

We also examine whether AGN photo-ionization is contributing to the high [NII]/H$\alpha$ ratios seen in some of our galaxies.  Figure~\ref{fig:emlines}$b$ shows the FMOS-COSMOS version of the BPT diagram at $z\sim1.6$.  Four galaxies with CO measurements have all key diagnostic lines detected by FMOS.  None of our galaxies fall in the region where strong AGN contribution is expected, above the line separating AGN from star-forming galaxies at $z\sim1.6$ \citep{Kewley2013}.  While PACS-819 approaches this line, it is not detected in X-rays with $Chandra$.  Nevertheless, we cannot rule out the presence of a low-to-moderate luminosity AGN, especially because its strong [NII] emission also drives this galaxy above the local MZ relation.   For reference, in Figure~\ref{fig:emlines}$b$, we plot the data for the larger star-forming  galaxy population from FMOS-COSMOS (grey data points).  It is worth highlighting that both PACS-830 and PACS-299 fall within the locus of the star-forming population, clearly offset towards a higher ionization state of the ISM  compared to local galaxies \citep{Steidel2014,Kartaltepe2015}.  

\section{Concluding remarks}

Our observations of the molecular CO gas content of seven galaxies well-above the MS at $z\sim1.6$ with ALMA in Cycle 1 and IRAM/PdBI establish an offset in the $L^{\prime}_{\rm CO}-L_{\rm TIR}$ relation for starbursts at high redshift, compared to MS galaxies, although smaller than expected from previous studies of low-z starburst outliers.  These results may be indicative of a continuous distribution in SFE at high redshift, as a function of distance from the star-forming MS, as opposed to a bi-modal distribution.  

An appealing physical explanation for a decrease in the L$^{\prime}_{\rm CO}$/L$_{\rm TIR}$ ratio for galaxies well-above the MS is galaxy mergers that can lead to rapid gas compression hence effectively boosting star formation  resulting in shorter gas depletion timescales.  Several galaxies well-above the MS are indeed in a merger phase \citep{Rodighiero2011,Wuyts2011}.  Within the present sample, hints of merging come from the presence of multiple UV/optical emitting regions  (e.g., in PACS-819, PACS-830, PACS-867), even being in a kinematically-linked system (PACS-164).  

\acknowledgments

We are grateful for the support from the regional ALMA ARCs, in particular Akiko Kawamura, Kazuya Saigo, Gaelle Dumas and Sergio Martin.  JDS was supported by the ALMA Japan Research Grant of NAOJ Chile Observatory, NAOJ-ALMA-0127.  This work was supported by World Premier International Research Center Initiative (WPI Initiative), MEXT, Japan. CM and AR acknowledge support from an INAF PRIN 2012 grant.  This paper makes use of the following ALMA data: ADS/JAO.ALMA\#2012.1.00952.S. ALMA is a partnership of ESO (representing its member states), NSF (USA) and NINS (Japan), together with NRC (Canada), NSC and ASIAA (Taiwan), and KASI (Republic of Korea), in cooperation with the Republic of Chile. The Joint ALMA Observatory is operated by ESO, AUI/NRAO and NAOJ.

\end{document}